\documentclass[twoside]{article}

\usepackage{amsmath,amssymb,amsthm,graphicx,enumerate,bm,natbib,epsfig,url}

\def\kj{{\mathcal{K}_j}}

\newcommand{\ind}[1]{\bm{1}\left({#1}\right)}
\usepackage[accepted]{aistats2e}
\begin{document}

%

%

\twocolumn[


\aistatstitle{Bayesian learning of joint distributions of objects}

\aistatsauthor{ Anjishnu Banerjee \And Jared Murray \And David B. Dunson }

\aistatsaddress{ Statistical Science, Duke University } ]

\begin{abstract}
 There is increasing interest in broad application areas in defining flexible joint models for data having a variety of measurement scales, while also allowing data of complex types, such as functions, images and documents. We consider a general framework for nonparametric Bayes joint modeling through mixture models that incorporate dependence across data types through a joint mixing measure.  The mixing measure is assigned a novel infinite tensor factorization (ITF) prior that allows flexible dependence in cluster allocation across data types.  The ITF prior is formulated as a tensor product of stick-breaking processes.  Focusing on a convenient special case corresponding to a Parafac factorization, we provide basic theory justifying the flexibility of the proposed prior and resulting asymptotic properties.  Focusing on ITF mixtures of product kernels, we develop a  new Gibbs sampling algorithm for routine implementation relying on slice sampling. The methods are compared with alternative joint mixture models based on Dirichlet processes and related approaches through simulations and real data applications.
\end{abstract}

\section{INTRODUCTION}

There has been considerable recent interest in joint modeling of data of widely disparate types, including not only real numbers, counts and categorical data but also more complex {\em objects}, such as functions, shapes, and images.  We refer to this general problem as {\em mixed domain modeling} (MDM), and major objectives include exploring dependence between the data types, co-clustering, and prediction. Until recently, the emphasis in the literature was almost entirely on parametric hierarchical models for joint modeling of mixed discrete and continuous data without considering more complex {\em object data}.  The two main strategies are to rely on underlying Gaussian variable models \citep{M84} or exponential family models, which incorporate shared latent variables in models for the different outcomes \citep{SR97,D00,D03}. Recently, there have been a number of articles using these models as building blocks in discrete mixture models relying on Dirichlet processes (DPs) or closely-related variants \citep{CS11,SX09,YD10}. DP mixtures for mixed domain modeling were also considered by \citet{HB11,SN09,BD10} among others. Related approaches are increasingly widely-used in broad machine learning applications, such as for joint modeling of images and captions \citep{LZ11}, and have rapidly become a standard tool for MDM.
  
Although such joint Dirichlet process mixture models (DPMs) are quite flexible, and can accommodate joint modeling with complicated objects such as functions \citep{BD09}, they suffer from a key disadvantage in relying on conditional independence given a single latent cluster index. For example, as motivated in \citet{D09,D10}, the DP and related approaches imply that two subjects $i$ and $i'$ are either allocated to the same cluster ($C_i=C_{i'}$) {\em globally} for all their parameters or are not clustered.  The soft probabilistic clustering of the DP is appealing in leading to substantial dimensionality reduction, but a single global cluster index conveys several substantial practical disadvantages.  Firstly, to realistically characterize joint distributions across many variables, it may be necessarily to introduce many clusters, degrading the performance in the absence of large sample sizes. Secondly, as the DP and the intrinsic Bayes penalty for model complexity both favor allocation to few clusters, one may over cluster and hence obscure important differences across individuals, leading to misleading inferences and poor predictions.  Often, the posterior for the clusters may be largely driven by certain components of the data, particularly when more data are available for those components, at the expense of poorly characterizing components for which less, or more variable, data are available.

To overcome these problems we propose Infinite Tensor Factorization (ITF) models, which can be viewed as {\em next generation} extensions of the DP to accommodate dependent object type-specific clustering. Instead of relying on a single unknown cluster index, we propose separate but dependent cluster indices for each of the data types whose joint distribution is given by a random probability tensor. We use this to build a general framework for hierarchical modeling. The other main contribution in this article is to develop a general extension of blocked sliced sampling, which allows for an efficient and straightforward algorithm for sampling from the posterior distributions arising with the ITF; with potential application in other multivariate settings with infinite tensors, without resorting to finite truncation of the infinitely many possible levels. 

\section{PRELIMINARIES}

We start by considering a simple bivariate setting $p=2$ in which data for subject $i$ consist of $y_i = (y_{i1},y_{i2})' \in \mathcal{Y}$, with $\mathcal{Y} =  \mathcal{Y}_1 \otimes \mathcal{Y}_2$, $y_{i1} \in \mathcal{Y}_1$, and $y_{i2} \in \mathcal{Y}_2$ for $i=1,\ldots,n$.  We desire a joint model in which $y_i \sim f$, with $f$ a probability measure characterizing the joint distribution.  In particular, letting $\mathcal{B}(\mathcal{Y})$ denote an appropriate sigma-algebra of subsets of $\mathcal{Y}$, $f$ assigns probability $f(B)$ to each $B \in \mathcal{B}(\mathcal{Y})$.  We assume $\mathcal{Y}$ is a measurable Polish space, as we would like to keep the domains $\mathcal{Y}_1$ and $\mathcal{Y}_2$ as general as possible to encompass not only subsets of Euclidean space and the set of natural numbers but also function spaces that may arise in modeling curves, surfaces, shapes and images.  In many cases, it is not at all straightforward to define a parametric joint measure, but there is typically a substantial literature suggesting various choices for the marginals $y_{i1} \sim f_1$ and $y_{i2} \sim f_2$ separately. 

If we only had data for the $j$th variable, $y_{ij}$, then one possible strategy is to use a mixture model in which  
\begin{eqnarray}
f_j( B ) = \int_{\Theta_j} \mathcal{K}_j( B; \theta_j)dP_j( \theta_j ),\quad B \in \mathcal{B}(\mathcal{Y}_j), 
\label{eq:1}
\end{eqnarray}
where $\mathcal{K}_j(\cdot; \theta_j)$ is a probability measure on $\{\mathcal{Y}_1,\mathcal{B}(\mathcal{Y}_1)\}$ indexed by parameters $\theta_j \in \Theta_j$, $\mathcal{K}_j$ obeys a parametric law (e.g., Gaussian), and $P_j$ is a probability measure over $\{ \Theta_j, \mathcal{B}(\Theta_j) \}$. A nonparametric Bayesian approach is obtained by treating $P_j$ as a random probability measure and choosing an appropriate prior.  By far the most common choice is the Dirichlet process \citep{F73}, which lets $P_j \sim DP (\alpha P_{0j})$.  Under the \citet{S94} stick-breaking representation, one then obtains,
\begin{eqnarray}
\nonumber f_j( B ) = \sum_{h=1}^{\infty} \pi_h \mathcal{K}_j( B; \theta_h^* ),\mbox{with}\\
 \pi_h= V_h \prod_{l<h} (1-V_l),\quad \theta_h^* \sim P_{0j}, \label{eq:2}
\end{eqnarray}
and $V_h \sim \mbox{Be}(1,\alpha)$, so that $f_j$ can be expressed as a discrete mixture.  This discrete mixture structure implies the following simple hierarchical representation, which is crucially used for efficient computation: 
\begin{eqnarray}
y_{ij} \sim  \mathcal{K}_j( \theta_{C_i}^* ),\quad \theta_h^* \sim P_{0j}, \quad \mbox{pr}(C_i = h) = \pi_h, 
\label{eq:3}
\end{eqnarray}
where $C_i$ is a cluster index for subject $i$.  The great success of this model is largely attributable to the {\em divide and conquer} structure in which one allocates subjects to clusters probabilistically, and then can treat the observations within each cluster as separate instantiations of a parametric model.  In addition, there is a literature showing appealing properties, such as minimax optimal adaptive rates of convergence for DPMs of Gaussians \citep{SG11,T11}. 

The standard approach to adapt expression (\ref{eq:1}) to accommodate mixed domain data is to simply let $f(B) = \int_{\Theta} \mathcal{K}( B;\theta)dP(\theta)$, for all $B \in \mathcal{B}(\mathcal{Y})$, where $\mathcal{K}(\cdot; \theta)$ is an appropriate joint probability measure over $\{\mathcal{Y},\mathcal{B}(\mathcal{Y})\}$ obeying a parametric law.  Choosing such a joint law is straightforward in simple cases.  For example, \citet{HB11} rely on a joint exponential  family distribution formulated via a sequence of generalized linear models.  However, in general settings, explicitly characterizing dependence within $\mathcal{K}(\cdot; \theta)$ is not at all straightforward and it becomes convenient to rely on a product measure \citep{BD10}:
\begin{eqnarray}
\mathcal{K}(B; \theta) = \prod_j \mathcal{K}(B_j; \theta_j),\quad B =
\bigotimes_{j=1}^p B_j,\quad B_j \in \mathcal{B}
(\mathcal{Y}_j). 
\label{eq:prodkern}
\end{eqnarray}
If we then choose $P \sim DP(\alpha P_0)$ with $P_0 = \bigotimes_{j=1}^p P_{0j}$, we obtain an identical hierarchical specification to (\ref{eq:3}), but with the elements of $y_i = \{ y_{ij} \}$ {\em conditionally independent} given the cluster allocation index $C_i$. 

As mentioned in \S $1$, this conditional independence assumption given a single latent class variable is the nemesis of the joint DPM approach. We consider more generally  a multivariate $C_i = (C_{i1},\ldots,C_{ip})^T \in \{1,\ldots, \infty\}^p$, with separate but dependant indices across the disparate data types. We let,
\begin{eqnarray}
\nonumber \mbox{pr}(C_{i1}=h_1,\ldots,C_{ip}=h_p) = \pi_{h_1\cdots h_p},\\
\mbox{with} \quad h_j =1,\ldots,\infty, j=1,\ldots,p, \label{eq:prC}
\end{eqnarray}
where $\pi = \{ \pi_{h_1\cdots h_p} \} \in \Pi_p^{\infty}$ is an infinite $p$-way {\em probability tensor} characterizing the joint probability mass function of the multivariate cluster indices. It remains to specify the prior for the probability tensor $\pi$, which is considered next in \S3.

\section{PROBABILISTIC TENSOR FACTORIZATIONS}

\subsection{PARAFAC Extension}

Suppose that $C_{ij} \in \{1,\ldots,d_j\}$, with $d_j$ the number of possible levels of the $j$th cluster index.  Then, assuming that $C_i$ are observed unordered categorical variables, \citet{DX09} proposed a probabilistic Parafac factorization of the tensor $\pi$: 
\begin{eqnarray} 
\pi = \sum_{h=1}^{k} \lambda_h \psi_h^{(1)} \otimes \cdots \otimes \psi_h^{(p)},
\label{eq:DX}
\end{eqnarray}
where $\lambda = \{ \lambda_h \}$ follows a stick-breaking process, $\psi_h^{(j)} = (\psi_{h1}^{(j)},\ldots,\psi_{hd_j}^{(j)})^T$ is a probability vector specific to component $h$ and outcome $j$, $\otimes$ denotes the outer product.

We focus primarily on  generalizations of the Parafac factorization to the case in which $C_i$ is unobserved and can take infinitely-many different levels. We let, 
\begin{eqnarray}
\pi_{c_1 \cdots c_p} & = & \mbox{pr}(C_1=c_1,\ldots, C_p=c_p) =
\sum_{h=1}^{\infty} \lambda_h \prod_{j=1}^p \psi_{hc_j}^{(j)} \nonumber \\
\lambda_h & = & V_h \prod_{l<h} (1-V_l),\quad V_h \sim \mbox{Be}(1,\alpha)
\nonumber \\
\psi_{hr}^{(j)} & = & U_{hr}^{(j)} \prod_{s<r} (1-U_{hs}^{(j)}),\quad
U_{hr}^{(j)} \sim \mbox{Be}(1,\beta_j), \label{eq:new}
\end{eqnarray}
A more compact notation for this factorization of the infinite probability tensor $\pi$ is,
\begin{eqnarray}
\pi = \sum_{h=1}^{\infty} \lambda_h \bigotimes_{j=1}^p \psi_h^{(j)},\\
\lambda \sim \mbox{Stick}(\alpha),\ \psi_h^{(j)} \sim \mbox{Stick}(\beta_j),
\label{eq:new2}
\end{eqnarray}
which takes the form of a stick-breaking mixture of outer products of stick-breaking processes.  This form is carefully chosen so that the elements of $\pi$ are stochastically larger in those cells having the smallest indices, with rapid decreases towards zero as one moves away from the upper right corner of the tensor.

It can be shown that tensors realizations from the ITF distribution are valid in the sense that they sum to $1$ with probability $1$. We can be flexible in terms where exactly these cluster indices occur in a hierarchical Bayesian model. Next in \S  3.2, we formulate a generic mixture model for MDM, where the ITF is used characterize the cluster indices of the parameters governing the distributions of the disparate data-types.

\subsection{Infinite Tensor Factorization Mixture}

Assume that for each individual $i$ we have a data ensemble $(y_{i1},\ldots,y_{ip}) \in \mathcal{Y}$ where $\mathcal{Y} = \bigotimes_{j=1}^{p}\mathcal{Y}_j$. Let $\mathcal{B}(\mathcal{Y})$ be the sigma algebra generated by the product sigma algebra $\mathcal{B}(\mathcal{Y}_1)\times\cdots\times \mathcal{B}(\mathcal{Y}_p)$. Consider any Borel set $B = \bigotimes_{j=1}^p B_j \in \mathcal{B}(\mathcal{Y})$. Given cluster indices $(C_{i1}=c_{i1},\ldots,C_{ip}=c_{ip})$, we assume that the ensemble components are independent with 
\begin{align}
 \label{eq:ITM1}
 & f( y_{i1} \in B_1, \ldots, y_{ip} \in B_p\, |\, C_{i1}=h_1,\ldots,C_{ip}=h_p ) \nonumber\\ &= \prod_{j=1}^{p} \mathcal{K}_j(B_j;\theta_{j,h_j}).
\end{align}
$\mathcal{K}_j(\cdot;\theta_{j,h})$ is an appropriate probability measure on $\{\mathcal{Y}_j, \mathcal{B}(\mathcal{Y}_j)\}$ as in equation (\ref{eq:1}). Marginalizing out the cluster indices, we obtain
\begin{align}
  \label{eq:ITM2}
  & f(y_{i1} \in B_1, \ldots, y_{ip} \in B_p) \nonumber \\ &= \sum_{h_1=1}^{\infty} \cdots \sum_{h_p=1}^{\infty} \pi_{h_1,\ldots, h_p} \prod_{j=1}^{p} \mathcal{K}_j(B_j;\theta_{j,h_j}),
\end{align}
 where $\pi_{h_1,\ldots,h_p}=\mbox{pr}(C_{i1}=h_{1},\ldots,C_{ip}=h_{p})$. We let $\pi \sim \mbox{\small{ITF}}(\alpha,\beta)$ and we call the resulting mixture model an infinite tensor factorization mixture, $f \sim \mbox{\small{ITM}}(\alpha,\beta)$. To complete the model specification, we let $\theta_{j,h_j} \sim P_{0j}$ independently as in (\ref{eq:2}).

\par{} The model $y_i \sim f$, $f \sim \mbox{\small{ITM}}(\alpha,\beta)$, can be equivalently expressed in hierarchical form as 
\begin{eqnarray}
y_{ij} & \sim&  \mathcal{K}_j( \theta_{ij}^* ),\ \theta_i^* = P \sum_{h_1=1}^{\infty} \cdots \sum_{h_p=1}^{\infty} \pi_{h_1,\ldots, h_p}\prod_{j=1}^p \delta_{\theta_{j,h_j}}, 
\nonumber \\
\pi & \sim & \mbox{\small{ITF}}(\alpha,\beta),\quad \theta_{j,h_j} \sim P_{0j}, \label{eq:ITM3}
\end{eqnarray}
Here, $P$ is a joint mixing measure across the different data types and is given a infinite tensor process prior, $P \sim \mbox{\small{ITP}}(\alpha,\beta,\bigotimes_{j=1}^{p}P_{0j})$.  Marginalizing out the random measure $P$, we obtain the same form as in (\ref{eq:ITM2}).  The proposed infinite tensor process prior provide a much more flexible generalization of existing priors for discrete random measures, such as the Dirichlet process or Pitman Yor process.

\section{POSTERIOR INFERENCE }

\subsection{Markov Chain Monte Carlo Sampling}

We propose a novel algorithm for efficient exact MCMC posterior inference in the ITM model, utilizing blocked and partially collapsed steps. We adapt ideas from \citet{W07,PR08} to derive slice sampling steps with label switching moves, entirely avoiding truncation approximations. Begin by defining the augmented joint likelihood for an observation $y_i$, cluster labels $c_i = (c_{i0}, c_{i1},\dots,c_{ip})$ and slice variables $u_i = (u_{i0}, u_{i1},\dots,u_{ip})$ as
\begin{align}
& p(y_{i}, c_{i}, u_{i} \mid \lambda, \Psi, \Theta)  \nonumber\\ &= \ind{u_{i0}<\lambda_{c_{i0}}}\prod_{j=1}^p\kj(y_{ij}; \theta^{(j)}_{c_{ij}})\ind{u_{ij}<\psi_{c_{i0}c_{ij}}^{(j)}}\label{model:slice}
\end{align}

It is straightforward to verify that on marginalizing $u_i$ the model is unchanged, but including $u_i$ induces full conditional distributions for the cluster indices with finite support. Let $m_{0h}=\sum_{i=1}^n\ind{c_{i0}=h}$ and $\mathcal{D}_0 = \{h: m_{0h}>0\}$. Similarly define $m_{jhk}=\sum_{i=1}^n \ind{c_{i0}=h}\ind{c_{ij}=k}$ and $\mathcal{D}_j = \{k: \sum_{h=1}^\infty m_{jhk}>0\}$, and let $k^*_j=\max(\mathcal{D}_j)$ for $0\leq j\leq p$. Define  $\mathcal{U}_0= \{u_{i0}: 1\leq i\leq n \}$, $\mathcal{C}_0= \{c_{i0}: 1\leq i\leq n \}$, $\mathcal{U}_1 = \{u_{ij}: 1\leq i\leq n, 1\leq j\leq p\}$ and $\mathcal{C}_1 = \{c_{ij}: 1\leq i\leq n, 1\leq j\leq p\}$. The superscript $(-i)$ denotes that the quantity is computed excluding observation $i$.

\begin{enumerate}

\item Block update $(\mathcal{U}_0, \lambda, \alpha)$
\begin{enumerate}
\item Sample $(\alpha\mid \mathcal{C}_0)$. Standard results \citep{A74} give
\[
p(\alpha\mid \mathcal{C}_0)\propto p(\alpha)\alpha^{\tilde{c}} \frac{\Gamma(\alpha)}{\Gamma(\alpha+n)}
\]
for $\tilde{c} = |\mathcal{D}_0|$ which can be sampled via Metropolis-Hastings or using auxiliary variables when $p(\alpha)$ is a mixture of Gamma distributions \citep{EW95}.
\item Sample $(\lambda\mid \alpha, \mathcal{C}_0)$ by drawing $V_h\sim Beta(1+m_{0h}, \alpha+\sum_{l=h+1}^{k_0^*}m_{0l})$ for $1\leq h \leq k_0^*$ and setting $\lambda_h = V_h\prod_{l<h}(1-V_l)$
\item Label switching moves:
\begin{enumerate}
\item From $\mathcal{D}_0$ choose two elements $h_1, h_2$ uniformly at random and change their labels with probability $\min(1, (\lambda_{h_1}/\lambda_{h_2})^{m_{0h_2}-m_{0h_1}})$ 
\item Sample a label $h$ uniformly from $1,2,\dots,k_0^*$ and propose to swap the labels $h, h+1$ and corresponding stick breaking weights $V_h, V_{h+1}$. Accept with probability $\min(1, a)$ where

\[
a=\left(\frac{k_0^*}{k_0^*+1}\right)^{\ind{h=k^*_0}}
\frac{
\left(1-V_h\right)^{m_{0(h+1)}}
}{
\left(1-V_{h+1}\right)^{m_{0h}}
}
\]
\end{enumerate}
\item Sample $(u_{i0}|c_{i0}, \lambda)\sim U(0, \lambda_{c_{i0}})$ independently for $1\leq i\leq n$
\end{enumerate}

\item Update $\mathcal{C}_0$.
From \eqref{model:slice} the relevant probabilities are
\begin{align}
& Pr(c_{i0} = h | u_i, c_i, \Psi, \lambda) \nonumber\\ &\propto \ind{u_{i0}<\lambda_{h}}\prod_{j=1}^p\ind{u_{ij}<\psi_{hc_{ij}}^{(j)}}
\end{align}
However, it is possible to obtain more efficient updates through \emph{partial collapsing}, which allows us to integrate over the lower level slice variables and $\Psi$ instead of conditioning on them. Then we have
\begin{align}
 Pr&(c_{i0}=k \mid u_{i0}, \mathcal{C}_1, \mathcal{C}_0^{(-i)}, \lambda) \propto \ind{u_{i0}<\lambda_h} \nonumber\\
 & \times \prod_{j=1}^p 
\frac{\left(1+m^{(-i)}_{jkc_{ij}}\right)
  \prod_{l<c_{ij}} \left( \beta_k^{(j)}+\sum_{s>l}+m^{(-i)}_{jks}\right)
 }{
 \prod_{l\leq c_{ij}}\left(1+\beta_k^{(j)}+\sum_{s\geq c_{ij}}+m^{(-i)}_{jks}\right)\label{eq:c0fc}
 }
\end{align}
To determine the support of \eqref{eq:c0fc} we need to ensure that $u^*_0=\min\{u_{i0}:1\leq i\leq n\}$ satisfies
$
u_0^* > 1-\sum_{l=1}^{k_0^*} \lambda_l.
$
If $\sum_{l=1}^{k_0^*} \lambda_l < 1-u^*_0$ then
draw additional stick breaking weights $V_{k_0^*+1},\dots,V_{k_0^*+d}$ independently from $Beta(1, \alpha)$ until $\sum_{l=1}^{k_0^*+d} \lambda_l > 1-u^*_0$, ensuring that $\sum_{l=k_0^*+d+1}^\infty \ind{u_{i0}<\lambda_{h}}=0$ for all $1\leq i\leq n$. 
%
Then the support of \eqref{eq:c0fc} is contained within $1,2,\dots,k_j^*+d$ and we can compute the normalizing constant exactly.

%
\item Block update $(\mathcal{U}_1, \Psi, \beta)$:
\begin{enumerate}
\item Update $(\beta_r^{(j)}\mid \{c_{ij}: c_{i0}=r\}, \mathcal{C}_0)$ for $1\leq j\leq p$, $1\leq r\leq k_0^*$. If the concentration parameter is shared across global clusters (that is, $\beta_r^{(j)} \equiv \beta^{(j)}$) then a straightforward conditional independence argument gives
\begin{align}
& p(\beta^{(j)} \mid \{c_{ij}: c_{i0}=r\}, \mathcal{C}_0) \nonumber\\ &\propto 
p(\beta^{(j)})
\prod_{r\in \mathcal{D}_0}
\left(\beta^{(j)}\right)^{\tilde{c}_{jr}}
\frac{\Gamma(\beta^{(j)})}{\Gamma(\beta^{(j)}+n_r)}
\end{align}

where $n_r = |\{i : c_{i0}=r\}|$ and $\tilde{c}_{jr} = |\{h:m_{jrh}>0\}|$. Note that terms with $n_r=1$ (corresponding to top-level singleton components) do not contribute, since $\beta^{(j)}{\Gamma(\beta^{(j)})} = {\Gamma(\beta^{(j)}+1)}$. The updating scheme of \citet{EW95} is simple to adapt here using $|\mathcal{D}_0|$ independent auxiliary variables.

\item For $r\in \mathcal{D}_0$ update $(\psi^{(j)}_{r} \mid \mathcal{C}_0, \mathcal{C}_1, \beta_r^{(j)})$ by drawing $U^{(j)}_{rh}\sim Beta(1+m_{jrh}, \beta_r^{(j)}+\sum_{l=h+1}^{k_0^*}m_{jlh})$ for $1\leq h\leq k^*_j$

\item Label switching moves: For $1\leq j\leq p$,
\begin{enumerate}
\item From $\mathcal{D}_j$ choose two elements $h_1, h_2$ uniformly at random and change their labels with probability $\min(1, a)$ where
\[
a=\prod_{h_0\in \mathcal{D}_0}
\left(
\frac{\psi^{(j)}_{h_0h_1}}
{\psi^{(j)}_{h_0h_2}}
\right)^{m_{jh_0h_2}-m_{jh_0h_1}}
\]

\item Sample a label $h$ uniformly from $1,2,\dots,k^*_j$ and propose to swap the labels $h, h+1$ and corresponding stick breaking weights. Accept with probability $min(1, a)$ where

\begin{align}
&a=\left(\frac{k_j^*}{k_j^*+1}\right)^{\ind{h=k^*_j}} \nonumber \\ & \times \prod_{h_0\in \mathcal{D}_0}
\frac{
\left(1-U^{(j)}_{rh}\right)^{m_{jr(h+1)}}
} {
\left(1-U^{(j)}_{r(h+1)}\right)^{m_{jrh}}
}
\end{align}

\end{enumerate}
\item Sample $(u_{ij}|c_i, \Psi)\sim U(0, \psi^{(m)}_{c_{i0}c_{ij}})$ independently for $1\leq j\leq p$, $1\leq i\leq n$.
\end{enumerate}

\item Update $\mathcal{C}_j$ for $1\leq j\leq p$ independently. We have

\begin{align}
& Pr(c_{ij}=k \mid y, \Theta,u_{ij}, c_{i0}, \Psi) \nonumber\\ &\propto \kj(y_{ij}; \theta^{(j)}_{k})\ind{u_{ij}<\psi_{c_{i0}k}^{(j)}}
\end{align}

As in step 2 we determine the support of the full conditional distribution as follows: Let $u_{j}^* = \min\{ u_{ij}:1\leq i\leq n \}$. For all $r\in \mathcal{D}_0$, if
$\sum_{h=1}^{k_{j}^{*}}\psi_{rh}^{(j)}<1-u_{j}^*$ then extend the stick breaking measure $\psi^{(j)}_r$ by drawing $d_r$ new stick breaking weights from the prior so that $\sum_{h=1}^{k_{j}^{*}+d_r}\psi_{rh}^{(j)}>1-u_{j}^*$. Draw $\theta^{(j)}_{k_{j}^{*}+1},\dots,\theta^{(j)}_{k_{j}^{*}+d}\sim p(\theta^{(j)})$ independently (where $d=\max\{d_r:r\in \mathcal{D}_j\}$). Then update $c_{ij}$ from

\begin{align}
&Pr(c_{ij}=k \mid y, \Theta,u_{ij}, c_{i0}, \Psi) \nonumber\\ &= \frac{
\kj(y_{ij}; \theta^{(j)}_{k})\ind{u_{ij}<\psi_{c_{i0}k}^{(j)}}
}{
\sum_{h=1}^{k_j^*+d} \kj(y_{ij}; \theta^{(j)}_{k})\ind{u_{ij}<\psi_{c_{i0}k}^{(j)}}
}
\end{align}

\item Update $(\Theta|-)$ by drawing from
\[
p(\theta_{h}^{(j)}\mid y, \mathcal{C}_j)\propto p(\theta_{h}^{(m)})\prod_{\{i:c_{ij}=h\}}\kj (y_{ij}; \theta^{(j)}_{h})
\]
for each $1\leq j\leq p$ and $1\leq h \leq k^*_j$
\end{enumerate}

\subsection{Inference}

Given samples from the MCMC scheme above we can estimate the predictive distribution as

\begin{align}
&\hat{f}(y_{n+1}\mid y_n)= \frac{1}{T}\sum_{t=1}^T\sum_{h_0=1}^{k^*_0}\sum_{h_1=1}^{k^*_1}\cdots \sum_{h_p=1}^{k^*_p} \lambda^{(t)}_{h_0} \nonumber \\ & \times \prod_{j=1}^p\psi^{(t)}_{h_{0}h_j}\kj\left(y_{(n+1)j};\theta^{(j)(t)}_{h_j}\right)\label{eq:predest}
\end{align}

Each of the inner sums in \eqref{eq:predest} is a truncation approximation, but it can be made arbitrarily precise by extending the stick breaking measures with draws from the prior and drawing corresponding atoms from $p(\theta^{(j)})$. In practice this usually isn't necessary as any error in the approximation is small relative to Monte Carlo error.

\par{} The other common inferential question of interest in the MDM settings is the dependence between components, for example testing whether component $j1$ and $j2$ are independent of each other. As already noted, the dependence between the components comes in through the dependence between the cluster allocations and therefore, tests for independence between $j1$ and $j2$ is equivalent to testing for independence between their latent cluster indicators  $C_{j1}$ and $C_{j2}$. Such a test can be constructed in terms of the divergence between the joint and marginal posterior distributions of $C_{j1}$ and $C_{j2}$. The Monte Carlo estimate of the Kulback Leibler divergence between the joint and marginal posterior distributions is given as,
\begin{align}
I&(j1,j2) =\frac{1}{T}\sum_{t=1}^{T} \sum_{h_{j1}=1}^{k^*_{j1}}\sum_{h_{j2}=1}^{k^*_{j2}} \left( \sum_{h_0=1}^{k^*_0} \lambda^{(t)}_{h_0} \psi^{(t)}_{h_{0}h_{j1}} \psi^{(t)}_{h_{0}h_{j2}} \right) \nonumber \\
 & \times \log \left( \frac{\sum_{h_0=1}^{k^*_0} \lambda^{(t)}_{h_0} \psi^{(t)}_{h_{0}h_{j1}} \psi^{(t)}_{h_{0}h_{j2}}}{\left[\sum_{h_0=1}^{k^*_0} \lambda^{(t)}_{h_0} \psi^{(t)}_{h_{0}h_{j1}}\right] \left[\sum_{h_0=1}^{k^*_0} \lambda^{(t)}_{h_0} \psi^{(t)}_{h_{0}h_{j1}}\right]}\right) 
\end{align}
Under independence, the divergence should be $0$. Analogous divergences can be considered for testing other general dependancies, like $3$-way, $4$-way independences. 

\section{EXPERIMENTS} 
 Our approach can be used for two different objectives in the context of mixed domain data - for prediction and for inference on the dependence structure between different data types.  We outline results of experiments with both simulated and real data that show the performance of our approach with respect to both the objectives.

\subsection{Simulated Data Examples} 

To the best of our knowledge, there is no standard model to jointly predict for mixed domain data as well as evaluate the dependence  structure, so as a competitor, we use a joint  DPM. To keep the evaluations fair, we use two scenarios. In the first the ground truth is close to that of the joint  DPM, in the sense that all the components of the mixed data have the same cluster structure. The other simulated experiment considers the case when the ground truth is close to the ITF, where different components of the mixed data ensemble have their own cluster structure but clustering is dependent. The goal here in each of the scenarios is to compare joint  DPM vs ITF in terms of recovery of dependence  structure and predictive accuracy. 

For scenario 1, we consider a set of 1,000 individuals from whom an ensemble comprising of T, a time series R, a multivariate real-valued response ($\in \Re^4$) and C1,C2,C3, 3 different categorical variables have been collected, to emulate the type of data collected from patients in cancer studies and other medical evaluations. For the purposes of scenario 1, we simulate T, R, C1, C2, C3 each from a mixture of 3 clusters. For example, R is simulated from a two-component mixture of multivariate normals with different means,  R is simulated from  a mixture of two autoregressive kernels  and each of the categorical variables from a mixture of two multinomial distributions. If we label the clusters as $1$ and $2$, for each simulation, either all of the ensemble (T,R,C1,C2,C3) comes from $1$ or all of it comes from $2$. After simulation we randomly hold out R in 50 individuals, C1, C2 in 10 each, for the purposes of measuring prediction accuracy. For the categorical variables prediction accuracy is considered with a $0-1$ loss function and is expressed as a percent missclassification rate. For the multivariate real variable R, we consider squared error loss and accuracy is expressed as relative predictive error. We also evaluate for some of the pairs their dependence  via estimated mutual information.

For scenario 2, the same set-up as in scenario 1 is used, except for the cluster structure of the ensemble. Now simulations are done such that T falls into three clusters and this is dependent on R and C1. C2 and C3 depend on each other  and are simulated from two clusters each but their clustering is independent of the other variables in the ensemble. We measure prediction accuracy using a hold out set of the same size as in scenario 1 and also evaluate the dependence  structure from the ITF model.

In each case, we take 100,000 iterations of the MCMC scheme with the first few 1,000 discarded as a burn-in.  These are reported in  table \ref{t1} (left). We also summarize the recovered dependence  structure in table \ref{t1}  and in table \ref{t2}. In scenario 1, the prediction accuracy of ITF and DPM are comparable, with DPM performing marginally better in a couple of cases. Note that the recovered dependence  structure with the ITF is exactly accurate which shows that the ITF can reduce to joint  co-clustering when that is the truth. In scenario 2, however there is significant improvement in using the ITF over the DPM with predictive accuracy. In fact the predictions from the DPM for the categorical variable are close to noise. The dependence  structure recovered the ITF almost reflects the truth as compared to that from the DPM which predicts every pair is dependent, by virtue of its construction.

\subsection{Real Data Examples}
For generic real mixed domain data the dependence  structure is wholly unknown. To evaluate how well the ITF does in capturing pairwise dependencies, we first consider a network example in which recovering dependencies is of principal interest and prediction is not relevant. We consider data comprising of 105 political blogs \citep{AG05} where the edges in the graph  are composed of the links between websites. Each blog is labeled with its ideology, and we also have the source(s)  which were used to determine this label.  Our model includes the network,  ideology label, and binary indicators for 7 labeling sources (including  ``manually  labeled'', which are thought to be the most subject to errors in labelings). We assume  that ideology impacts  links through cluster  assignment only,  which is a reasonable  assumption here. We collect 100,000 MCMC  iterations after  a short burn-in  and  save the  iterate  with  the  largest  complete-data likelihood for exploratory purposes.  

\begin{figure}[h]
\vspace{.3 in}
\begin{minipage}[b]{.75\linewidth}
\centering
\includegraphics[width=\textwidth]{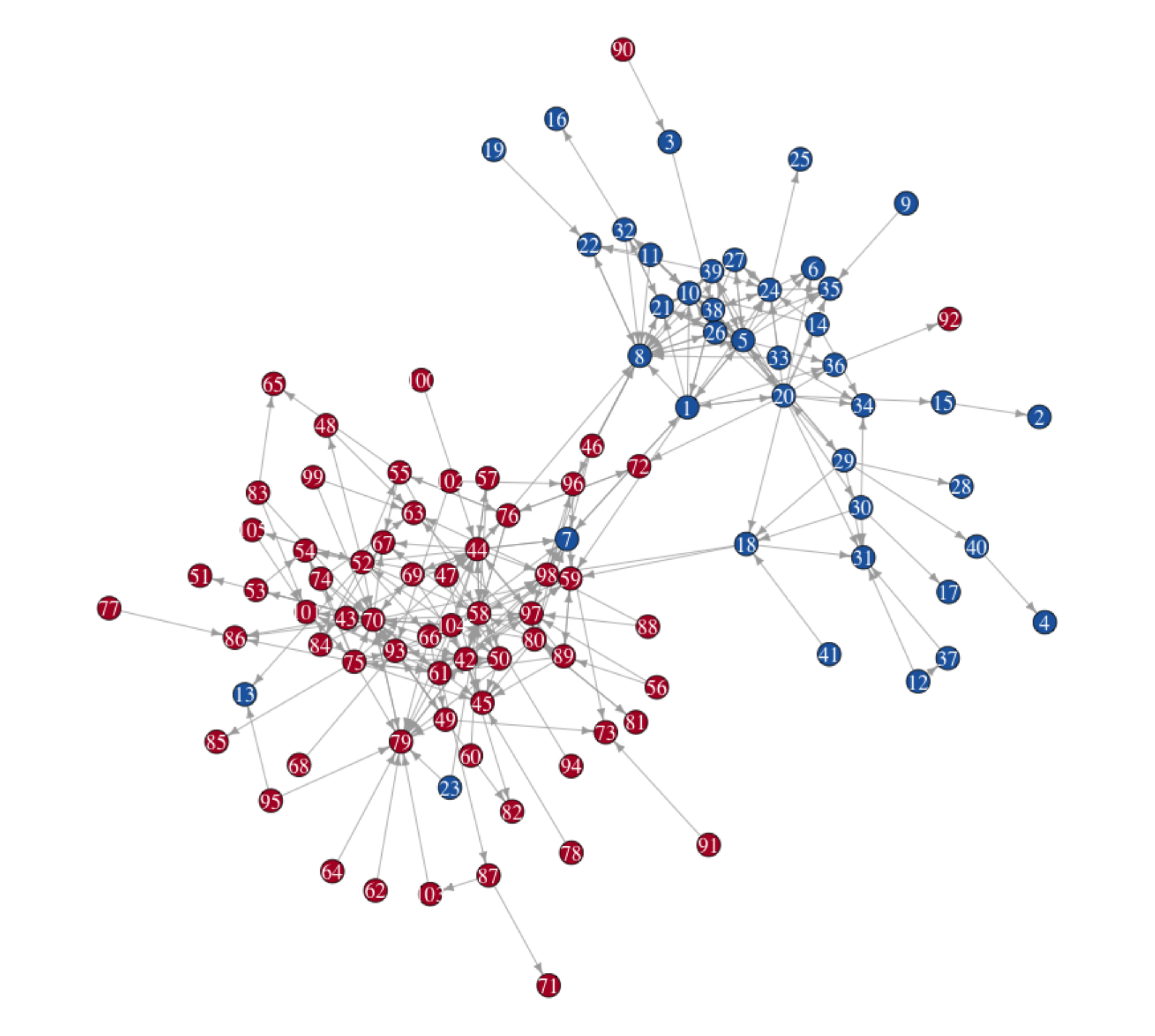}
\vspace{.3 in}
\caption{Network Example: True Clustering}
\vspace{.3 in}
\label{f1}
\end{minipage}
\hspace{0.5cm}
\begin{minipage}[b]{.75\linewidth}
\centering
\includegraphics[width=\textwidth]{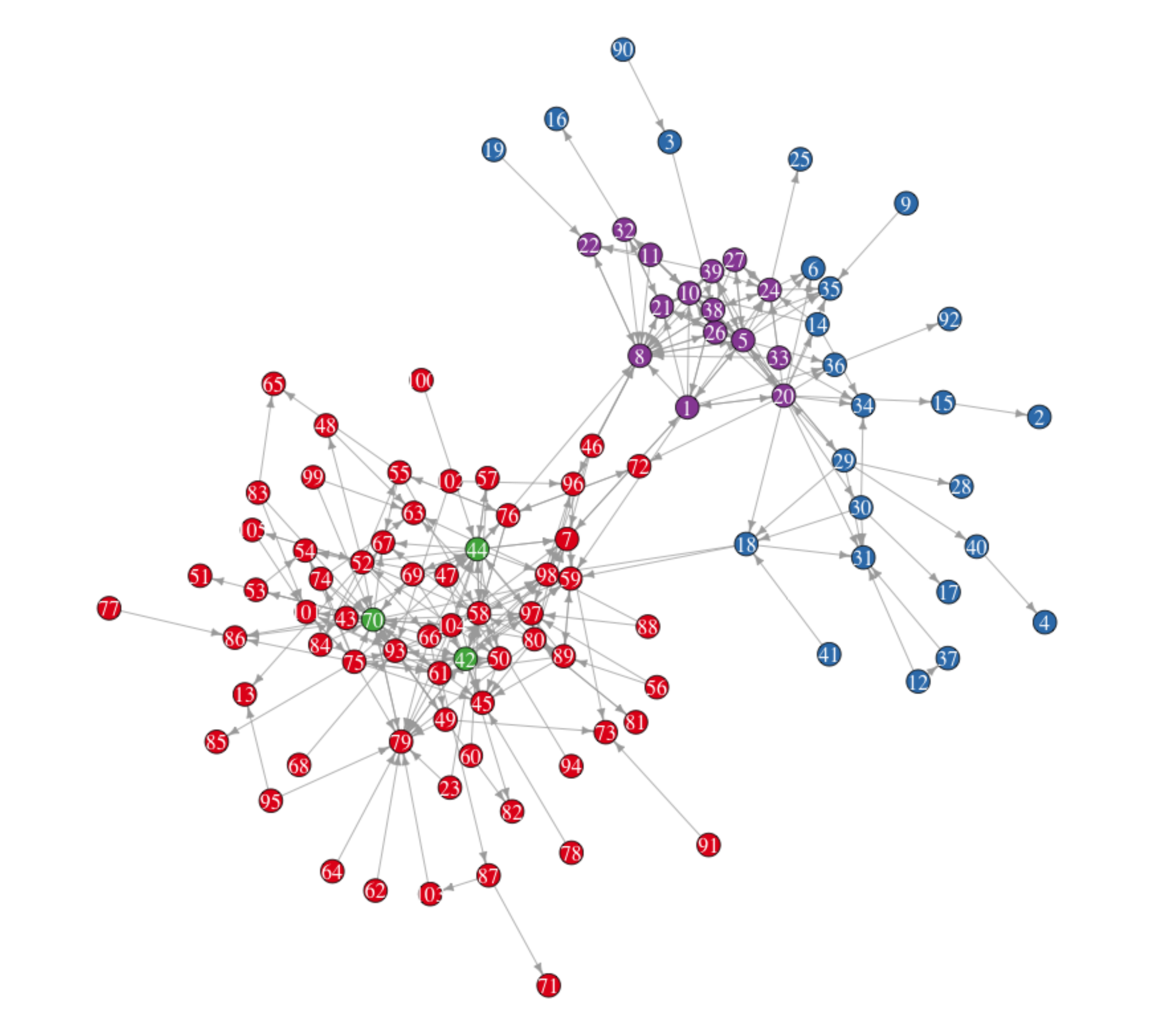}
\caption{Network Example: Recovered Clustering}
\label{f2}
\end{minipage}
\end{figure}

Fig. \ref{f1} shows the  network  structure, with  nodes colored by ideology.  It is immediately  clear that there  is significant  clustering,  apparently driven  largely by ideology, but  that ideology alone does not  account  for all the  structure present  in the  graph. Joint DPM approach would allow for only one type of clustering and prevent us from exploring this additional structure. The  recovered  clustering  in fig. \ref{f2} reveals a number  of interesting structural properties  of the  graph;  for example,  we see a tight cluster  of conservative  blogs which  have  high  in- and  out-  degrees  but  do not  link to  one another  (green) and a partitioning of the liberal blogs into a tightly  connected  component (purple) and a periphery component with low degree (blue).  The conservative  blogs do not exhibit  the same level of assortative mixing (propensity to link within a cluster)  as the liberal blogs do, especially within the purple component.

To get a sense for how stable  the clustering  is, we estimate  the posterior  probability that nodes $i$ and $j$ are assigned to the same cluster  by recording  the number  of times this event occurs in the MCMC. We observe that the clusters are generally quite stable,  with two notable  exceptions.  First, there is significant posterior  probability that points 90 and 92 are assigned to the red cluster  rather than  the blue cluster.  This is significant because these two points are the conservative  blogs which are  connected  only to  liberal  blogs (see fig. \ref{f1}).   While  the  graph  topology  strongly  suggests  that these  belong to the  blue cluster,  the  labels are able to exert  some influence as well.  Note that we do not observe the same phenomenon for points 7, 15, and 25, which are better connected.  We also observe some ambiguity  between  the  purple  and  blue clusters.   These  are nodes 6, 14, 22, 33, 35 and 36, which appear at the intersection of the purple/blue clusters in the graph projection  because they  are not  quite  as connected  as the  purple  ``core''  but  better connected  than  most  of the  blue clusters. 

\begin{figure}[h]
\hspace{0.5cm}
\vspace{.3 in}
\begin{minipage}[b]{.75\linewidth}
\centering
\includegraphics[width=\textwidth]{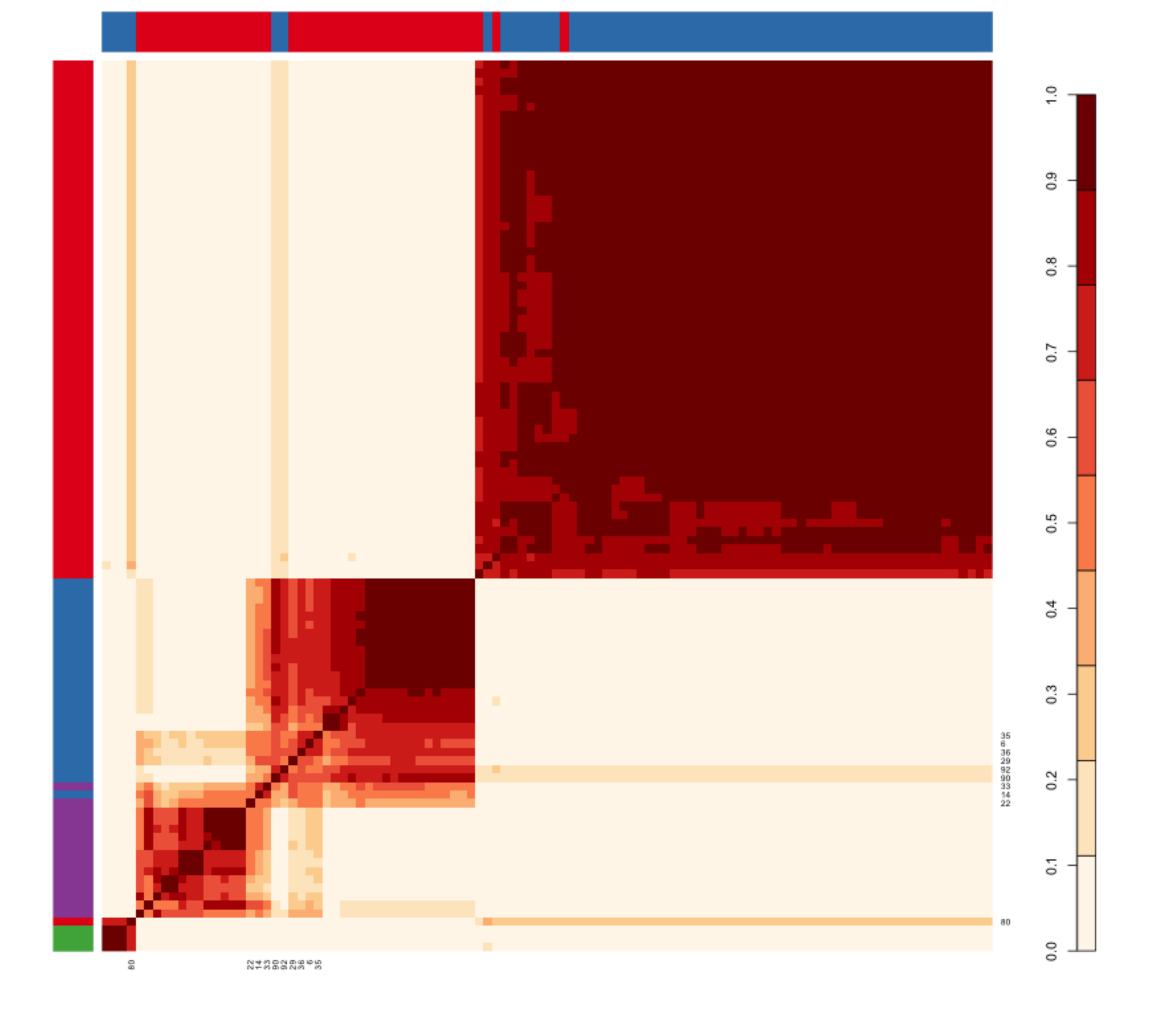}
\vspace{.3 in}
\caption{Network Example: Pairwise cluster assignment probability. Left bars correspond to clustering  in Fig. \ref{f2}, top bars correspond  to clustering  on the ideology label.}
\label{f4}
\end{minipage}
\end{figure}

Finally,  we examine  the  posterior  probability of being labeled "conservative" (fig. \ref{f4}).   Most data points are assigned very high or low probability. The  five labeled points  stand  out  as having uncharacteristic labels for their  link structure (see fig \ref{f1}).  Since the  observed label doesn't  agree with the graph  topology, the probability is pulled away from 0/1  toward a more conservative  value. This  effect is most  pronounced  in the  three  better-connected liberal  blogs (lower left)  versus the weakly connected  conservative  blogs (upper  right).

For the second example, we use data obtained from the Osteoarthritis Initiative (OAI) database, which is available for public access at \url{http://www.oai.ucsf.edu/}. The question of interest for this data is investigate relationships between physical activity and knee disease symptoms. For this example we use a subset of the baseline clinical data, version 0.2.2. The data ensemble comprises of variables including biomarkers, knee joint symptoms, medical history, nutrition, physical exam and subject characteristics. In our subset we take an ensemble of size $120$ for $4750$ individuals. We hold out some of the biomarkers and knee joint symptoms and consider prediction accuracy of the ITF versus the joint  DPM model. For the real variables, mixtures of normal kernels are considered, for the categorical, mixtures of multinomials and for the time series, mixtures of fixed finite wavelet basis expansion. 

Results for this experiment are summarized in table \ref{t3} for 4 held-out variables. ITF outperforms the DPM in 3 of these 4 cases and marginally worse prediction accuracy in case of the other variable. It is also interesting to note that ITF helps to uncover useful relationships between medical history, physical activity and knee disease symptoms, which has a potential application for clinical action and treatments for the subsequent patient visits.

\section{CONCLUSIONS}

We have developed a general model to accommodate complex ensembles of data, along with a novel algorithm to sample from the posterior distributions arising from the model. Theoretically, extension to any number of levels of stick breaking processes should be possible, the utility and computational feasibility of such extensions is being studied. Also under investigation is connections with random graph/network models and theoretical rates of posterior convergence.

\begin{table}[h]
\caption{Simulation Example, Scenario 1: Prediction error (top), tests of independence (bottom)}
\label{t1}
\begin{minipage}[b]{0.9\linewidth}\centering
\begin{tabular}{|c|c|c|}
&ITF&DPM\\
\hline
T& 1.79 & 1.43 \\
C2& 31$\%$ & 23 $\%$\\
C3& 37$\%$ & 36 $\%$\\
\hline
\end{tabular}
\end{minipage}
\hspace{0.5cm}
\begin{minipage}[b]{0.9\linewidth}
\centering
\begin{tabular}{|c|c|c|c|}
&ITF&DPM&``Truth''\\
\hline
C1 vs T & Yes & Yes & Yes\\
C2 vs T & Yes & Yes & Yes\\
C3 vs T & Yes & Yes & Yes\\
C2 vs R & Yes & Yes & Yes\\
\hline
\end{tabular}
\end{minipage}
\end{table}

\begin{table}[h]
\caption{Simulation Example, Scenario 2: Prediction error (top), tests of independence (bottom)}
\label{t2}
\begin{minipage}[b]{0.9\linewidth}\centering
\begin{tabular}{|c|c|c|}
&ITF&DPM\\
\hline
T& 4.61 & 10.82 \\
C2& 27$\%$ & 55 $\%$\\
C3& 34$\%$ & 57 $\%$\\
\hline
\end{tabular}
\end{minipage}
\hspace{0.5cm}
\begin{minipage}[b]{0.9\linewidth}
\centering
\begin{tabular}{|c|c|c|c|}
&ITF&DPM&``Truth''\\
\hline
C1 vs T & Yes & Yes & Yes\\
C2 vs T & No & Yes & No\\
C3 vs T & No & Yes & No\\
C2 vs R & No & Yes & No\\
\hline
\end{tabular}
\end{minipage}
\end{table}

\begin{table}[h]
\caption{OAI Data example: Relative Predictive Accuracy. The variables are respectively, left knee baseline pain, isometric strength left knee extension,left knee paired X ray reading, left knee baseline radiographic OA.}
\label{t3}
\begin{minipage}[b]{0.75\linewidth}\centering
\begin{tabular}{|c|c|c|}
&ITF&DPM\\
\hline
P01BL12SXL & 31.21 & 100.92 \\
V00LEXWHY1& 7.94 & 7.56 $\%$\\
V00XRCHML& 23.01 & 31.84 $\%$\\
P01LXRKOA & 65.78 & 90.30 $\%$\\
\hline
\end{tabular}
\end{minipage}
\vspace{.3 in}
\end{table}

\subsubsection*{Acknowledgements}

This work was support by Award Number R01ES017436 from the National Institute of Environmental Health Sciences and DARPA MSEE.  The content is solely the responsibility of the authors and does not necessarily represent the official views of the National Institute of Environmental Health Sciences or the National Institutes of Health or DARPA MSEE.

\bibliographystyle{biometrika}
\bibliography{mybib}

\end{document}